\documentclass[prl,aps,twocolumn,amsmath,amssymb,
showpacs,preprintnumbers,superscriptaddress]{revtex4-1}
\usepackage{color,graphicx,tikz,
dcolumn,bm}

\definecolor{grau}{gray}{.5}
\definecolor{schwarz}{gray}{0}

\newcommand{\reff}[1]{(\ref{#1})}

\newcommand{\av}[1]{\left|#1\right|}

\newcommand{\brkts}[1]{\left(#1\right)}

\newcommand{\bsplitl}[2]{
\begin{equation}
\begin{split}
#1
\end{split}
\label{#2}
\end{equation}}
\newcommand{\bsplit}[1]{
\begin{equation*}
\begin{split}
#1
\end{split}
\end{equation*}}

\begin{document}
\title[A new stochastic mode reduction strategy]{A new
stochastic mode reduction strategy for dissipative systems}

\author{M. Schmuck}%
\affiliation{Department of Chemical Engineering,	Imperial College London, London SW7 2AZ, United Kingdom}
\affiliation{Department of Mathematics,	Imperial College London, London SW7 2AZ, United Kingdom}
\author{M. Pradas}%
\affiliation{Department of Chemical Engineering, Imperial College London, London SW7 2AZ, United Kingdom}
\author{S. Kalliadasis}%
\affiliation{Department of Chemical Engineering, Imperial College London, London SW7 2AZ, United Kingdom}
\author{G. A. Pavliotis}%
\affiliation{Department of Mathematics, Imperial College London, London SW7 2AZ, United Kingdom}
\date{\today}%

\begin{abstract}
We present a new methodology for studying non-Hamiltonian nonlinear systems
based on an information theoretic extension of a renormalization group
technique using a modified maximum entropy principle. We obtain a rigorous
dimensionally reduced description for such systems. The neglected
degrees of freedom by this reduction are replaced by a systematically defined
stochastic process under a constraint on the second moment. This then forms
the basis of a computationally efficient method.
Numerical computations for the generalized Kuramoto-Sivashinsky equation
support our method and reveal that the long-time underlying stochastic
process of the fast (unresolved) modes obeys a universal distribution which
does not depend on the initial conditions and which we rigorously derive by
the maximum entropy principle.
\end{abstract}

 \maketitle

Many nonlinear time dependent problems in science and engineering are too
complex to be fully resolved and hence some degrees of freedom need to be
neglected. Popular examples of high dimensional problems, which can only be
solved or studied by model reduction and approximation, are models for weather and
climate prediction, cell biology processes, non-linear networks,  or economics.
These problems involve many different timescales, e.g.~oceans show a characteristic
behavior over years whereas atmosphere over days. More generally, any process governed by
nonlinear partial differential equations (PDEs) is infinite dimensional and hence
requires a reliable finite dimensional representation for numerical purposes. This rises
the central question of how can one systematically and reliably reduce the complexity of
such high-dimensional systems without neglecting essential information contained in
the unresolved/neglected degrees of freedom.

As in the case  of weather and climate modelling, time-scale separation is 
a central feature of  many dissipative processes in physical and
industrial applications. Such a scale separation can be conveniently
expressed in Fourier space  by differentiating the so-called fast  modes
which are characterized with large wavenumbers and converge towards
equilibrium much faster than the  slow (low wavenumber) modes. As a
consequence, the long-time behavior of the system is primarily contained in
the latter. There are  several mode reduction techniques in the
literature that take advantadge of such scale decomposition, including
deterministic methodologies such as adiabatic
elimination~\cite{VanKampen1985}, and the classical center-manifold
theory~\cite{Foias1988,*Roberts90,*Fujimura91,*Gallay93}, which usually
requires the system to be close to criticality and to equilibrium, i.e. close
to an invariant solution which is expected to be part of a finite dimensional
center manifold~\cite{HaragusBook}. A different line of thought follow the
so-called stochastic mode reduction strategies, where the aim is to convert
an infinite-dimensional deterministic dynamical system (PDE) into a
low-dimensional stochastic one. In the case of Hamiltonian-like systems,
there exist well-known powerful techniques such as optimal prediction and the
Mori-Zwanzig formalism~\cite{Mori1965,*Zwanzig1961,*Zwanzig1973} for the
derivation of mode reduced (low-dimensional) stochastic equations. Both
approaches make use of the
existence of an invariant or canonical probability distribution given by the Hamiltonian structure. 
Other examples of these include \cite{Majda2002} where a Galerkin truncated Burgers-Hopf equation is considered
to obtain a Hamiltonian structure and a canonical Gibbs measure.

However, there does not seem to exist a stochastic methodology
available for problems which do not have a Hamiltonian structure. It is
precisely our purpose to study this open problem. To this end, we properly
extend the \emph{evolutionary renormalization group (ERG)
method}~\cite{Moise2001,*Temam1999}, which asymptotically defines equations
for slow and fast modes, towards a stochastic mode reduction by adapting the
\emph{principle of maximum information entropy (PMIE)}~\cite{Jaynes1957,
*Lasota1994,*Shannon1948} appropriately extended to PDEs, which allows
to extract the relevant information from the fast modes
to obtain a closed equation for the slow modes only. 
Separation of scales (slow/fast) can be physically justified by the presence of
dissipation. We observe that the ERG method provides a systematic and rigorous (in
terms of error estimates) tool to separate a dissipative, nonlinear problem
into fast ($w^\epsilon$) and slow ($v^\epsilon$) modes: 
\bsplitl{
\partial_t v^\epsilon
	& = f(v^\epsilon,w^\epsilon)\,,
\qquad
\partial_t w^\epsilon
	= \frac{1}{\epsilon}g(v^\epsilon,w^\epsilon)\,,
}{FastSlow}
where the parameter $0<\epsilon\ll 1$ measures the timescale separation. 
It is important to note that in the aforementioned deterministic mode
reduction strategies such as invariant manifolds, one generally assumes that
the model of interest decomposes as \reff{FastSlow} from the beginning
without further specification $\epsilon \ll 1$. However, such a
decomposition and the associated definition of $\epsilon$ needs to be checked
carefully from a practical and theoretical point of view, and in particular
for systems exhibiting spatio-temporal chaos, it is expected that many
characteristic spatio-temporal scales can be present. A well-known example of
such systems is the generalized Kuramoto-Sivashinsky  (gKS)
equation~\cite{Tseluiko2010,*Tseluiko2012} which retains the fundamental
ingredients of any nonlinear process involving spatio-temporal transitions
and pattern formation: nonlinearity, instability/energy production,
stability/energy dissipation and dispersion. In our 
methodology, we systematically define the parameter $\epsilon$, which
mediates the time scales and depends on the number of the resolved degrees of
freedom represented by $v^\epsilon$. We can further place our low dimensional
approximation on a solid theoretical basis by rigorous error estimates
\cite{Schmuck2011} controled by $\epsilon$. Moreover, our
stochastic mode reduction based on the PMIE rigorously supports theoretically
Stinis' computational approach~\cite{Stinis}. Using the dissipation property
defined by the renormalized fast modes, which implies that the fast Fourier
modes are independently distributed, we show via PMIE that the they are
Gaussian distributed with zero mean and variance $\sigma^2>0$. This leads to
a rigorous methodology to rationally and systematically derive stochastic
low-dimensional representations of deterministic, nonlinear, and
non-Hamiltonian PDEs  and further explain the arising randomness as the
extracted information from the neglected fast modes,  which is important
from the point of view of noise-induced
phenomena~\cite{Pradas_PRL,*Pradas_EJAM,Hutt_PRL07,*Hutt_PhysD08,*Hutt_EurPhysLett12}.
Finally, our new stochastic mode reduction strategy rigorously supports the
formal RG approach applied to the gKS equation by Chow and
Hwa~\cite{CHOW1995}.

\paragraph{General methodology: ERG and PMIE.--}
Consider dissipative nonlinear PDEs of the form
\bsplitl{
\partial_t u
	= {\cal A}u + {\cal F}(u)\,,
}{NlPDE}
where $u(x,t)$ is a one-dimensional variable with space and time
dependence,  ${\cal A}$ denotes a linear spatial
differential operator and ${\cal F}$ a nonlinear term. For the ease
of presentation we consider deterministic initial conditions (ICs)
with high enough spatial regularity (i.e.~differentiability), and
for simplicity, we restrict ourselves to periodic boundary conditions. 
The ERG method consists in approximating $u:=v+w$ by 
$u^\epsilon:=v^\epsilon+w^\epsilon$, where $v^\epsilon$ are the slow 
and $w^\epsilon$ the fast modes, i.e.,
\bsplitl{
\partial_t v^\epsilon
	& = {\cal A}_vv^\epsilon + {\rm P}_N{\cal F}(v^\epsilon,w^\epsilon)\,,
\\
\partial_t w^\epsilon
	& = {\cal A}_ww^\epsilon + {\rm Q}_N{\cal F}(v^\epsilon,w^\epsilon)\,,
}{RGFaSl}
where ${\rm P}_N$ and ${\rm Q}_N:= I-{\rm P}_N$ are the orthogonal
projections to the first $N$ Fourier modes and its complement. We have also
applied formally the notation ${\cal A}_v:={\rm P}_N{\cal A}$ and ${\cal
A}_w:= \epsilon{\rm Q}_N{\cal A}$ where $\epsilon:=\frac{1}{N^\beta}$ and 
$\beta>0$ denotes the highest order of spatial derivatives defined by the 
operator ${\cal A}$. The separation \reff{RGFaSl} can be made rigorous
by error estimates~\cite{Schmuck2011} which indicate how large one should 
choose $N>0$. Next,  insert the asymptotic expansion
$u^\epsilon=u^0+\epsilon u^1 +\epsilon^2 u^2+\dots$ into \reff{RGFaSl}
reformulated for a vectorial $u^\epsilon:=[v^\epsilon,w^\epsilon]'$, 
leading to the ERG equation
\begin{equation}\label{RGeq}
\partial_t U
	= {\cal F}_R(U),
\end{equation}
with $U(0)=u_0$ which removes secular terms growing in time  (see
e.g.~\cite{Schmuck2011cms,*Schmuck2012jmp} for classical homogenization 
with respect to space). $U$ is  the  RG solution which turns out to
be the Galerkin approximation with $N$ Fourier basis functions, and also
decomposes into slow $V$ and fast $W$ modes.  The resonant part ${\cal F}_R$
of ${\cal F}$ is defined via 
\bsplitl{ {\rm e}^{L\tau}{\cal
F}\brkts{{\rm e}^{-L\tau}u_0}-{\cal A}v_0 =: {\cal F}_R(u_0)+\tilde{\cal
F}_{NR}(\tau, u_0)\,, }{FR}
where $L$ is the system size, $\tilde{\cal F}_{NR}$ represents the
non-resonant part, and $\tau$ is the rescaled time. The slow variable $V$ of
the RG equation \reff{RGeq} solves the standard Galerkin approximation of
\reff{NlPDE} for $2N+1$ modes, i.e., 
\bsplitl{ \partial_t V
	= {\cal A}_v V
	+ {\rm P}_N{\cal F}(V)\,.
}{Glrkn}
Putting things together finally
leads to the renormalized solutions
\bsplitl{
{\rm v}^\epsilon
	& = {\rm P}_N u^\epsilon
	= V(t)+\epsilon {\rm PF}_{NR}(t/\epsilon,U)\,,
\\
{\rm w}^\epsilon
	& = {\rm Q}_N u^\epsilon
	= {\rm e}^{-{\rm Q}_N{\cal A}t/\epsilon}(W(t)+\epsilon {\rm QF}_{NR}(t/\epsilon,U))\,,
}{RnSl}
where ${\rm F}_{NR}(s,U):=\int^s_0\tilde{\cal F}_{NR}(\tau,U)\,d\tau$. Note that
the fast modes $W$ (required in \reff{RGeq}) are still infinite dimensional. 
To obtain a finite-dimensional representation, we   replace ${\rm w}^\epsilon$ 
with a random process which is defined by the original (i.e., not renormalized) 
fast variable $w^\epsilon$ which contains more information about the dynamics.
We obtain the probability distribution for  $w^\epsilon$ via PMIE by
maximizing the information entropy
\bsplitl{
{\cal S}_I(f(w^\epsilon_j))
	:= -\int_\Omega f(w^\epsilon_j){\rm log}\brkts{\frac{f(w^\epsilon_j)}{\nu(\omega)}}\,d\omega\,,
}{InfEntr}
under the constraint
\bsplitl{
\int_\Omega f(w^\epsilon_j)\frac{{\rm d}}{{\rm d} t} {\cal C}_N(w^\epsilon_j)\,d\omega
	= \delta_j(t)
	\,,
}{Cnstrnt}
where $\Omega$ is the space of events and $f(w^\epsilon_j)$ is the
probability density of the $j$-th fast mode $w^\epsilon_j(t,\omega)$;
$\delta_j(t)$ is a characteristic dissipation rate, which for simplicity we
approximate by the $j$-th Fourier mode
$\tilde{W}_j(t):={\rm e}^{-\rho_j^wt/\epsilon}w_j(0)$ of the leading order 
term in \reff{RnSl}$_2$ by setting $\delta_j(t):=\frac{{\rm d}}{{\rm d}t}{\cal
C}_N(\tilde{W}_j):=-\frac{1}{2}\frac{{\rm d}}{{\rm d}t}\tilde{W}_j^2$.
The measure $\nu$ is defined by prior or
background knowledge on the system, such as uncertainties associated with
the model  (which turns out to be a uniform distribution after applying the PMIE), and
the stochastic process $W$ defined via \reff{InfEntr} and \reff{Cnstrnt} finally
leads to a random process for the 
solution ${\rm v}^\epsilon$ in \reff{RnSl}$_1$.
%
\paragraph{The gKS equation.--}We exemplify the above procedure with
the gKS equation, i.e.,
\bsplit{
{\cal A}
	& := -\brkts{
\partial_x^2
	+ \kappa\partial_x^3
	+ \partial_x^4
	}
	\,,
\quad{\cal F}(u)
	:= -u\partial_x u\,,
}
defined on the periodic domain ${\cal D}_L:=]-L/2,L/2[$.
The ERG provides the deterministic approximation \reff{RnSl} via \reff{Glrkn} and
\bsplitl{
&
{\rm PF}_{NR}(s,U)
	=2i\lambda \sum_{\av{j}\leq N}{\rm e}^{i\frac{j}{\alpha}x}
		\sum_{\substack{
				k+l=j\\
				\av{k}\leq N <\av{l}
			}
		} \frac{{\rm e}^{-\rho_l^ws}}{\rho_l^w}V_k\frac{j}{\alpha}W_l
\\&\qquad\quad
		+i\lambda\sum_{\av{j}\leq N}{\rm e}^{i\frac{j}{\alpha}x}
		\sum_{\substack{
				k+l=j\\
				\av{k},\av{l}>N
			}		
		} \frac{{\rm e}^{-(\rho_k^w+\rho_l^w)s}}{\rho_k^w+\rho_l^w}W_k\frac{j}{\alpha}W_l \nonumber
		\,,
}{PFnr}
where
$\rho_l^w:=-\frac{1}{N^\beta}\brkts{\brkts{l/\alpha}^2+i\kappa\brkts{l/\alpha}^3+\brkts{l/\alpha}^4}$
are the eigenvalues of ${\cal A}_w$ with 
eigenvectors ${\rm
e}^{i\frac{k}{\alpha}x}$, and $\alpha:=L/2\pi$.

We choose $ {\cal C}_N(w^\epsilon_j(t,\omega)):=-\frac{1}{2}(w^\epsilon_j)^2(t,\omega)$
and hence from the solutions of the 
equation for $w^\epsilon_j$, we obtain
\bsplitl{
\frac{{\rm d}}{{\rm d} t}{\cal C}_N(w^\epsilon_j)
=\rho_j^w(w^\epsilon_j)^2+iw^\epsilon_j\sum_{\substack{\av{k}\leq N\\ \av{j-k}>N}}
v_k\frac{j-k}{\alpha}w^\epsilon_{j-k}
\, \nonumber .
}{Cwj}
%
Maximizing now the information entropy \reff{InfEntr} under the constraint
\reff{Cnstrnt}, see \cite[Chapt. 9]{Lasota1994}, leads to the following probability density
$f(w^\epsilon_j)
	:= \frac{1}{Z_j}m_j\frac{1}{\sigma_j\sqrt{2\pi}}\exp{-\frac{(w^\epsilon_j-\mu_j)^2}{2\sigma_j^2}}\,,	
$ 
where $m_j:=c_j^{-1}\sigma_j\sqrt{2\pi}\exp{-\mu_j^2/(2\sigma_j^2)}$, and
\bsplitl{
\mu_j
	:= \frac{i}{2\rho^w_j}\sum_{\substack{|k|\leq N\\ |j-k|>N}}v_k^\epsilon\frac{j-k}{\alpha}\mu_{j-k}
	\,,
\textrm{ }
\sigma_j^2
	:= \frac{1}{2\lambda_j\rho^w_j}\,.
}{sigmu}
Note that if one replaces \reff{Cnstrnt} by $\int_{-\infty}^\infty
f(w^\epsilon_j)(w^\epsilon_j)^2\,d\omega=\sigma^2$,  we get the
classical result $w_j\sim {\cal N}(0,\sigma^2)$. We apply the PMIE 
with respect to the original dynamics $w^\epsilon$ in \reff{InfEntr} 
and hence added complete dynamical information by to the Fourier 
modes $W_j$ of $W$ in \reff{sigmu} leading to 
$W_j\sim{\cal N}(\mu_j,\sigma_j^2)$. It follows from \reff{sigmu} and
$\overline{\mu_j}=\mu_{-j}$,  where $\overline{\mu}_j$ denotes the
complex conjugate of $\mu_j$, that $\mu_j=0$ for $|j|>N$ and the constraint
\reff{Cnstrnt} finally defines $\lambda_j=1/(2\delta_j(t))$ where
$\delta_j(t)=\frac{{\rm d}}{{\rm d}t}{\cal C}_N(\tilde{W}_j)
=\rho_j^w{\rm e}^{-2\rho_j^wt}w_j^2(0)$ for the dissipation rate in 
\reff{Cnstrnt}. For random ICs, the above methodology carries over 
but $w_k^2(0)$ is replaced with $\langle w_k^2(0) \rangle$, where 
brackets denote average over different ICs.

From the formula for $f(w^\epsilon_j)$ we deduce that 
the normalization constant, i.e.~the partition function 
is $Z_j:= m_j$. Based on $f$, the distribution of the fast renormalized 
variable $W(x,t):=\sum_{|j|>N}W_j{\rm e}^{ij/\alpha(x+2 V_0t)}$ can be 
determined as $W\sim {\cal N}(\mu_W,\sigma_W^2)$ where
%
$\mu_W
	:= \sum_{|j|>N} {\rm e}^{ij/\alpha (x+2 V_0t)}\mu_j$, and
$\sigma_W^2
	:= \sum_{|j|>N}{\rm e}^{i2j/\alpha (x+2 V_0t)}\sigma_j^2$.
%
Using $W\sim{\cal N}(\mu_W,\sigma_W^2)$ in equation \reff{RnSl}$_1$ gives the
final result of our stochastic mode reduction method. Not only this result
offers a systematic way of  accurately representing deterministic equations
with low-dimensional stochastic ones but it also allows for efficient
computations as we only need to solve the reduced model and  add the noise
\emph{a posteriori}.

\begin{figure}
\centering
\includegraphics[width=0.45\textwidth]{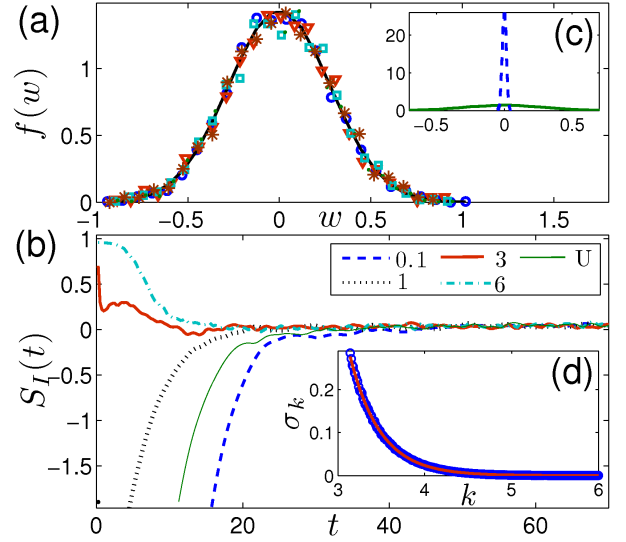}
\caption{(Color online) (a) Long-time behaviour of
$f(w_k)$ for the first fast mode ($k=N+1$) and five different initial
conditions ( represented by different symbols). The solid line is a Gaussian PDF with
$\mu=0$ and $\sigma=\sigma_{N+1}=0.27$. The inlet (c) shows snapshots of $f(w)$
at $t=0$  (dashed line) and long-times (solid line). (b) Time evolution
of the entropy $S_I$ for $k=N+1$ and random initial conditions
of the form $u_0(x)=a\xi(x)$ with $a=1$ for a uniform distribution (U),
and $a=0.1$, $1$, $3$, and $6$ for a Gaussian distribution. For simplicity
we have considered $\nu(w)=1$ to compute $S_I$ in \reff{InfEntr}.  (d)
Exponential $k$-dependency of $\sigma$ at long times where
the solid line is a data fit with 
$\sigma_k=\exp{(-2.86k+7.73)}$.  $k$ is rescaled here by $\alpha=L/2\pi$. }\label{Fig: S stat}
\end{figure}%

\paragraph{Numerical results and physical interpretations.--}We start by 
looking at the statistics of the fast modes by numerically solving  the gKS
equation for $2\Lambda+1$ Fourier modes with $\Lambda=2048$,  
$\kappa=0.1$, and using different types of random ICs of the form
$u_0(x)=a\xi(x)$, where $\xi(x)$ corresponds to either spatial white noise
[i.e.~$\langle\xi(x)\xi(x')\rangle=2\delta(x-x')$] with zero mean and unit
variance or a uniform distribution $\xi(x)\in [-1,1]$.  We choose also different
values for the  noise amplitude, namely $a=0.1,\,1,\,3,\,6$, and perform $2,000$
noise realizations each. The spatio-temporal solution of $u(x,t)$ rapidly 
evolves into a complex dynamics characterised by a chaotic behaviour
(see e.g.~\cite{Tseluiko2010,*Tseluiko2012}).
Figures \ref{Fig: S stat}(a,c) show that after some time, the distribution of 
the fast modes relaxes to a universal PDF  which is
independent of the ICs and corresponds to a Gaussian
distribution ${\cal N}(0,\sigma_k^2)$.  This relaxation  can
also be seen by computing the evolution of the entropy $S_I$,
observing that the final state is independent of the ICs
[cf.~\ref{Fig: S stat}(b)].  Our results also suggest that the
variance of the fast modes has a $k$-dependency which is 
an exponential decay as  a consequence of dissipation.
\begin{figure}
\centering
\includegraphics[width=0.47\textwidth]{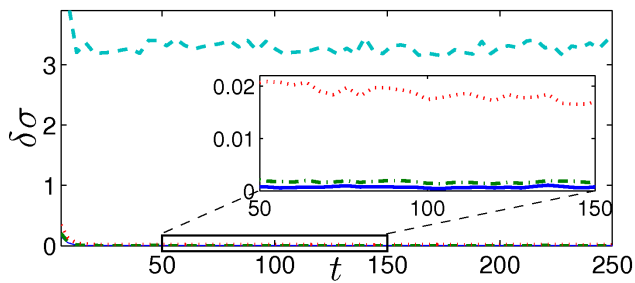}
\caption{(Color online) Difference between the second moment
of the full and ERG solutions as function of time and for
$N=768$ (cyan dashed line),
$896$ (red dotted line), $960$ (green dot-dashed line) and $1024$ (blue solid line).
The inlet shows a zoom into the marked area.}\label{Fig: Dsigma}
\end{figure}%

Next, we demonstrate the performance of our new mode reduction strategy by
comparing numerical results obtained from the full system, i.e.~the solution $u(x,t)$
that takes into account all the modes up to $\Lambda$, and the reduced 
system, i.e.~the solution  $v^\epsilon(x,t)$ obtained with a lower number $N<\Lambda$ 
of modes. 
We note that the efficiency of the method relies on the fact that we only
need to solve the system with $N<\Lambda$ modes and add the ERG corrections
\emph{a posteriori} [see Eq.~\reff{RnSl}] where the fast modes  $W_k$ are
given as random variables of zero mean and variance $\sigma_k$, the
$k$-dependency of which is the one observed numerically in Fig.~\ref{Fig: S
stat}. We compute first the second moment of both the full spatio-temporal
solution, i.e.~$\sigma_F=(\overline{u^2}-\overline{u}^2)^{1/2}$, where the
overline denotes spatial average, and the ERG solution $v^\epsilon(x,t)$,
which we denote as $\sigma_R$. Figure \ref{Fig: Dsigma} depicts the error
distance between both magnitudes: $\delta\sigma:=\langle
(\sigma_F-\sigma_R)^2\rangle$, for different values of $N$. We observe that
the ERG solution converges to the full solution (with $\delta\sigma<
10^{-3}$) as $N\to\Lambda/2$. 
%
\begin{figure}
\centering
\includegraphics[width=0.48\textwidth]{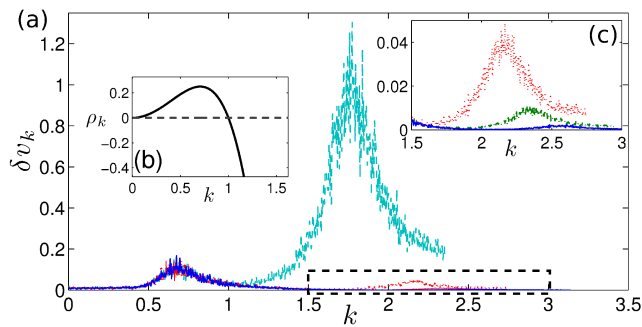}
\caption{(Color online) Difference between the modes obtained
from the full and ERG solution by taking  $N=768$ (cyan dashed line),
$896$ (red dotted line), $960$ (green dot-dashed line) and $1024$ (blue solid line).
 $k$ is rescaled here by $\alpha=L/2\pi$ and the maximum values correspond to
$k_\mathrm{max}=2.35$, $2.74$, $2.94$, and $3.13$, respectively. 
(b) Eigenvalues $\rho_k$ of the linear
operator ${\cal A}$. (c) Zoom into the area
marked with dashed lines.}\label{Fig: vk}
\end{figure}%

We also compute the error distance between each Fourier mode by defining:
$\delta v_k=\langle ((v_k-v_k^\epsilon)/\mathrm{max}(v_k)))^2\rangle$, 
where $v_k$ and $v_k^\epsilon$ correspond to the absolute value  (amplitude) of the 
slow modes from the full and ERG solution.
Figure \ref{Fig: vk} shows the results for sufficiently long times
(i.e.~the entropy function has already equilibrated)
and for different values of $N$. It is interesting to note that the error distance 
for the linearly unstable (slow) modes, which are defined as those
with positive eigenvalues $\rho_k>0$ [see Fig.~\ref{Fig: vk}(b)] is
practically not affected as we change $N$. On the other hand, the stable
modes between $1<k<3$ largely depend on the truncation number $N$ observing
as before a rapid convergence to zero as $N\to\Lambda/2$. This indicates that
(a) the dynamics of the unstable modes is robust and seems not to depend on
the number of stable modes used for the reduced model, and (b) less than half
of the stable modes, i.e., $N<\Lambda/2$, provides already a reliable
representation of the full system solution.

%
Finally, we also look at how time-correlations of a single mode can be well
represented by the reduced model. To this end, we compute the frequency power
spectrum of the absolute value of a given mode for $0<t<T$,
i.e.~$s_f(k,w)=(1/T)\sum_{t}v_k(t)\exp{(iwt)}$ (See Supplemental Material at 
[]). 
As before, all solutions with different $N$ give similar results for the
unstable modes, whereas the difference between both
solutions for the stable slow modes grows as $N$ is decreased 
(see Fig.~1 in Supplemental Material at  []).

To conclude, we have outlined a new stochastic mode reduction methodology for
dissipative dynamical systems of the general form \reff{NlPDE}.  It was
exemplified with a paradigm for nonlinear evolution and pattern formation,
the gKS equation. The cornerstone of our methodology is the information
entropy which combined with an ERG formalism gives a rigorous and systematic justification
of fast-slow scale separation as well as of randomness in dissipative systems. We demonstrated numerically the validity of the method and its
efficiency, i.e.~that one only needs to solve the reduced model (which
can contain as few as half of the whole number of modes) and then add the
particular type of the underlying stochastic process resulting from the
maximum entropy principle. We further showed that the methodology allows to
uncover new physical insights. These include: a universal PDF for the fast
modes that emerges independently of the ICs and a clear distinction between
the  modes which are relevant to describe the dynamics of the full system
based on a reduced model, from those that have a faster decay.  Moreover,
our method uncovers an appropriate definition of entropy for dissipative
non-equilibrium processes which show a universal characteristic such as 
a Gaussian PDF, thus providing a systematic means for quantifying the
evolution of dissipative systems.

\acknowledgments We acknowledge financial support from EPSRC
grant No.~EP/H034587, EU-FP7 ITN via Grant No.~214919 (Multiflow) and
ERC via Advanced Grant No.~247031.

\bibliographystyle{apsrev4-1} 
\bibliography{stoMoReByRG72}
\end{document}